\documentclass[mypaper,8pt,twoside]{CoAst}
\usepackage{epsf,graphicx,fancyhdr}
\renewcommand{\chaptermark}[1]{\markboth{#1}{}}

\newcommand{\kopf}{\small\itshape Comm. in Asteroseismology \\ Contribution to the Proceedings of the Wroclaw HELAS Workshop, 2008}

\newcommand{\Authors}[1]{\begin{center}\normalsize\bf\sf #1 \end{center}}

\renewcommand{\author}[1]{\begin{center}\normalsize\bf\sf #1 \end{center}}
\newcommand{\Address}[1]{\begin{center}\small\sf #1 \end{center}}

\newcommand{\Session}[1]{{\vspace{3mm}\small \noindent  \hspace*{3mm} Session: } #1 \normalsize}

\newcommand{\Objects}[1]{{\vspace{0mm}\small \noindent  \hspace*{3mm} Individual Objects: } \small #1 \normalsize}

	\newcommand{\poster}{\small Poster \newline}

\renewenvironment{abstract}{\section*{Abstract}\normalsize\sf}{}
\newcommand{\References}[1]{\begin{flushleft}{\large References\\}\vspace*{2mm}\small #1 \end{flushleft}}

\newcommand{\chapterCoAst}[2]{\chapter[\sf\normalsize #1\\ \footnotesize \hspace*{5mm}by #2 \sf\normalsize][]{#1\\}\rhead[\fancyplain{}{\sf\footnotesize \center{#1}}]{\fancyplain{}{\sffamily\thepage}}\lhead[\fancyplain{\kopf}{\sffamily\thepage}]{\fancyplain{\kopf}{\sf\footnotesize \center{#2}}}}

\newcommand{\figureDSSN}[5]{\begin{figure}[#4]
\centering
\includegraphics*[#5]{#1}
\caption{#2}
\label{#3}
\end{figure}}

\renewcommand{\chaptername}{}

\newcommand{\acknowledgments}[1]{\vspace*{5mm}\noindent  \textbf{Acknowledgments.} #1}

\def\rfr{\smallskip\par\noindent
        \hangindent=7truemm
        \hangafter=1}

\begin{document}
\sf

\chapterCoAst{Solar-like oscillations in red giants in the CoRoT exo-field}
{S.\,Hekker, C.\,Barban, T.\,Kallinger, W.\,Weiss, J.\,De Ridder, A.\,Hatzes and the CoRoT team} 
\Authors{S.\,Hekker$^{1,2}$, C.\,Barban$^3$, T.\,Kallinger$^4$, W.\,Weiss$^4$, J.\,De Ridder$^2$, A.\,Hatzes$^5$ and the CoRoT team} 
\Address{
$^1$ Royal Observatory of Belgium, Ringlaan 3, 1180 Brussels, Belgium\\
$^2$ Instituut voor Sterrenkunde, KU Leuven, Celestijnenlaan 200D, 3001 Leuven, Belgium\\
$^3$ Observatoire de Paris, LESIA, CNRS UMR 8109, Place Jules Janssen, 92195 Meudon, France\\
$^4$ Institut f\"ur Astronomie, T\"urkenschanzstrasse 17, A-1180 Vienna, Austria\\
$^5$ Th\"uringer Landessternwarte Tautenburg, Sternwarte 5, 07778 Tautenburg, Germany\\
}

\noindent
\begin{abstract}
Asteroseismic observations from space can provide us with long time series of uninterrupted high quality data for many stars at the same time. The CoRoT satellite (Convection Rotation and planetary Transits) was launched successfully in December 2006 and provides high precision photometery for a large number of stars. Here we present our research on (late G and K) red giant stars observed  during the first long run (150 days) of CoRoT with the 'eye' dedicated to exo-planet research.  
\end{abstract}

\Session{ \poster } 
\Objects{red giant stars} 

\section*{Introduction}
Solar-like oscillations in red-giant stars are observed using both spectroscopy (e.g. Frandsen et al. 2002,  Barban et al. 2004, De Ridder et al. 2006) and photometry (e.g. Barban et al. 2007, Stello et al. 2007, Kallinger et al. 2008, Tarrant et al. 2008). Results from these observations left room for discussion about the observability of non-radial modes in red giants and the lifetime of the stochastic oscillations. Longer time series of data, such as the 150 day time series from CoRoT, could reveal more details about these phenomena. Here we present the selection and first analysis steps of red giants observed in the CoRoT exo-field during the first long run in the direction of the centre of the Milky Way (LRc01).

\section*{Selection}
In the CoRoT exo-field $\sim$ 12\,000 stars in the magnitude range 11-16 mag in V are observed during each run with a cadence of 512 or 32 seconds. Although the main focus was on red dwarfs, many red giants are observed. A selection procedure to identify the red giant candidates was developed based on amplitude spectra, using the following criteria: (1) excess amplitude in the frequency range 20 - 120 $\mu$Hz; (2) width of the excess of at least 20 $\mu$Hz or several peaks over a similar range; (3) some excess at low frequencies, which is interpreted as granulation. After an automatic selection based on the above criteria, all pre-selected stars are examined manually. From this selection it became clear that solar-like oscillations in this frequency regime are not observable for stars fainter than 15 mag in V, most likely due to photon noise. Finally, about 400 red giant candidates are selected. An example of a power spectrum of a red giant candidate is shown in Figure 1.

\section*{Data analysis}
Although CoRoT delivers high quality data, non-stellar trends and jumps are present in the light curves and need to be corrected. Therefore, a correction using the following steps is performed: (1) subtract a trend modelled by a $2^{\rm nd}$ order polynomial; (2) detect jumps by comparing mean flux values in adjacent time bins with a width of at least a few times the oscillation period; (3) fit polynomials in all parts separated by the jumps and divide the data by these polynomials. 

Currently a full analysis of the selected candidates is performed in different groups in Belgium, Paris, Tautenburg and Vienna. This includes the search for oscillation frequencies, mode identification, and lifetimes. Details of these analyses and results will be published in subsequent publications.

\figureDSSN{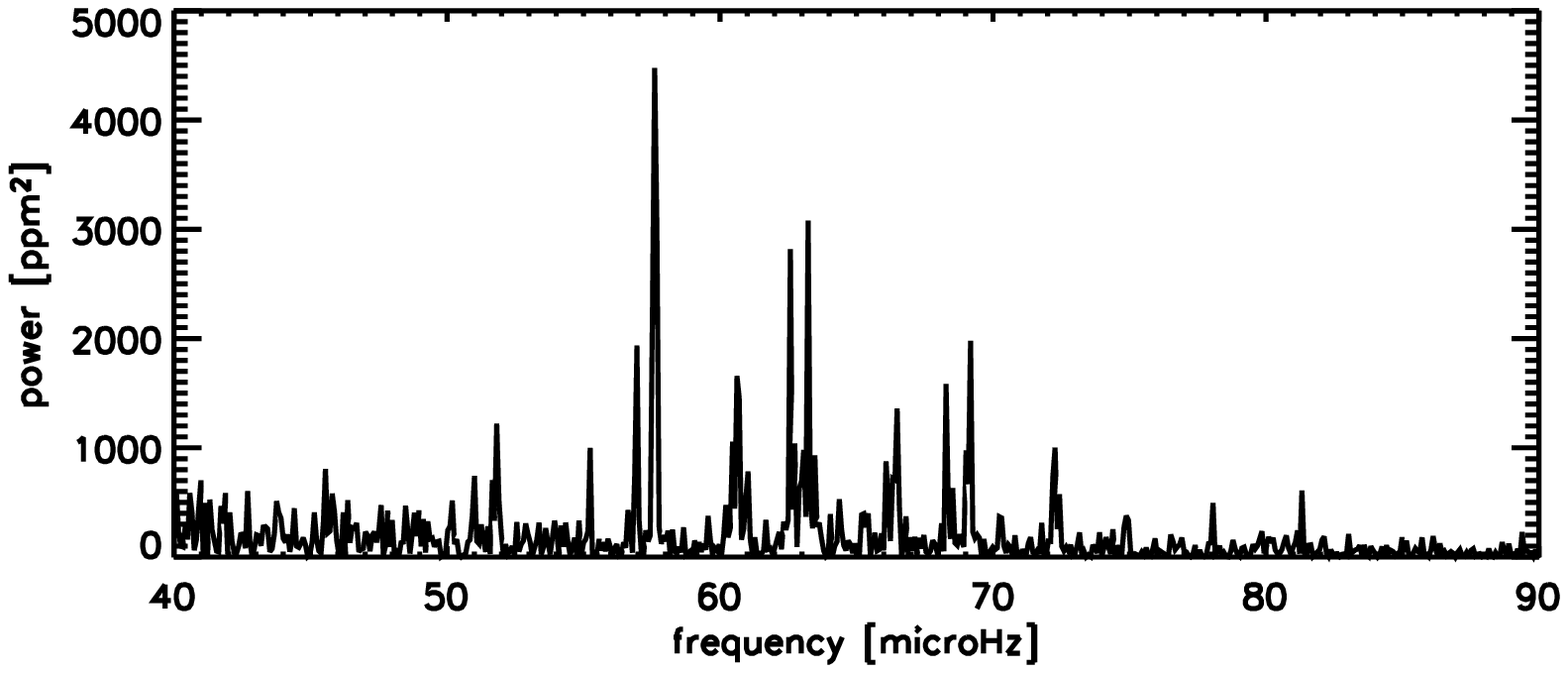}{Power spectrum of one of the selected red giant candidates. This star has an apparent magnitude of 12.5 mag in V, while no other parameters are available at this moment.}{label}{t}{clip,angle=0,width=120mm}


\acknowledgments{The CoRoT space mission, launched on December 27$^{\rm th}$ 2006, has been developed and is operated by CNES, with the contribution of Austria, Belgium, Brasil, ESA, Germany and Spain. SH acknowledges financial support from the Belgian Federal Science Policy (ref: MO/33/018). TK and WW are supported by the Austrian Science Funds P17580 and P7890. JDR is a postdoctoral fellow of the Fund of Scientific Research, Flanders.}

\References{
\rfr Barban C., De Ridder J., Mazumdar A. et al. 2004, ESA-SP, 559, 113
\rfr Barban C., Matthews J.M., De Ridder J. et al. 2007, A\&A, 468, 3, 1033
\rfr De Ridder J., Barban C., Carrier F. et al. 2006, A\&A, 448, 2, 689
\rfr Frandsen S., Carrier F., Aerts C. et al. 2002, A\&A, 394, L5
\rfr Kallinger T., Guenther D.B., Matthews J.M. et al. 2008, A\&A, 478, 2, 497
\rfr Stello D., Bruntt H., Kjeldsen H. et al. 2007, MNRAS, 377, 2, 584
\rfr Tarrant, N.J., Chaplin W.J., Elsworth Y. et al. 2008, A\&A, 483, 3, L43
}

\end{document}